\documentclass[amsmath,amssymb,aps,pre,twocolumn,groupedaddress]{revtex4-2}

\usepackage{graphicx}% Include figure files
\usepackage{dcolumn}% Align table columns on decimal point
\usepackage{bm}% bold math
\usepackage{hyperref}
\hypersetup{colorlinks=true, citecolor=blue, urlcolor=blue, linkcolor=blue}
\usepackage{epstopdf}
\begin{document}

%\preprint{}

\title{Thermodynamic efficiency of atmospheric motion governed by Lorenz system}

% author. The explanatory text should go in []'s, actual e-mail.

% \affiliation command. The affiliation command should follow the
\author{Zhen Li}
\email{li-zhen@g.ecc.u-tokyo.ac.jp}
\affiliation{Department of Complexity Science and Engineering, Graduate School of Frontier Sciences, The University of Tokyo, Kashiwa 277-8561, Japan.}
%\thanks{}

\author{Yuki Izumida}
\email{izumida@k.u-tokyo.ac.jp}
\affiliation{Department of Complexity Science and Engineering, Graduate School of Frontier Sciences, The University of Tokyo, Kashiwa 277-8561, Japan.}


%\date{\today}

\begin{abstract}
  {The Lorenz system was derived on the basis of a model of convective atmospheric motions and may serve as a paradigmatic model for considering a complex climate system.
In this study, we formulated the thermodynamic efficiency of convective atmospheric motions governed by the Lorenz system by treating it  {as} a non-equilibrium thermodynamic system. 
Based on the fluid conservation equations under the Oberbeck--Boussinesq approximation,
the work necessary to maintain atmospheric motion and heat fluxes at the boundaries were calculated. 
Using these calculations, the thermodynamic efficiency was formulated for stationary and chaotic dynamics.
The numerical results show that, for both stationary and chaotic dynamics, the efficiency tends to increase as the atmospheric motion is driven out of thermodynamic equilibrium when the Rayleigh number increases. However, it is shown that the efficiency is upper bounded by the maximum efficiency, which is expressed in terms of the parameters characterizing the fluid and the convective system.
The analysis of the entropy generation rate was also performed for elucidating the difference between the thermodynamic efficiency of conventional heat engines and the present atmospheric heat engine.
It is also found that there exists an abrupt drop in efficiency at the critical Hopf bifurcation point, where the dynamics change from stationary to chaotic. These properties are similar to those found previously in Malkus--Lorenz waterwheel system.}
\end{abstract}

%\keywords{}

\maketitle

\section{Introduction}
  {Understanding climate systems is of increasing importance in coping with climate change; the utility of physics approaches in this regard has been demonstrated~\cite{RevModPhys.92.035002,RevModPhys.94.015001,Lucarini2014}.}

  {The climate system can be regarded as a nonequilibrium thermodynamic system, which exchanges the energy and entropy with the surroundings~\cite{Lucarini2014,lucarini2010thermodynamics,RevModPhys.92.035002,RevModPhys.94.015001,PhysRevE.80.021118,EntropyBudgetOfTheatmosphere}.
Modeling the climate system as a heat engine driven by temperature differences has thus offered useful viewpoints~\cite{tritton2012physical,RevModPhys.94.015001, lorenz1955available}.
Heat-engine-analogs were proposed
by appropriately defining the heat inputs and outputs as well as the temperatures of hot and cold heat reservoirs in the climate system~\cite{johnson2000entropy,lucarini2010thermodynamics,EntropyProductionandClimateEfficiency,TowardQuantifyingtheClimateHeatEngineSolarAbsorptionandTerrestrialEmissionTemperaturesandMaterialEntropyProduction}. 
The impacts of global warming on climate thermodynamics, including thermodynamic efficiency, were studied using an Earth-like climate model with the variation of the $\text{CO}_2$ concentration~\cite{lucarini2010thermodynamics}, while the global entropy generation rate is often regarded as both a climate diagnostic and predictor~\cite{gibbins2020entropy}.
}

  {Understanding atmospheric motion is an important element in comprehending complex climate systems.
Convective atmospheric motions are caused by thermal imbalances owing to the difference in absorbed solar heat between upper and lower layers of the atmosphere.
They can be interpreted as the result of mechanical work dissipated by viscosity and used as a way to decrease thermal imbalances by convective heat transport~\cite{RevModPhys.94.015001, lorenz1967nature, peixoto1991entropy}.
Historically, Saltzman  {developed} a simplified model of time-dependent convective atmospheric motions and primarily solved it using a Fourier expansion~\cite{saltzman1962finite}. Based on Saltzman's model, Lorenz derived a set of nonlinear ordinary differential equations, known as Lorenz equations, which describe the motion of a specific mode of atmosphere, and discussed the stability of solutions~\cite{lorenz1963deterministic}.
The Lorenz equations are well known for yielding chaotic solutions under certain parameters and initial conditions, which may impact on the long-range weather forecasts~\cite{lorenz1963deterministic}.
Meanwhile, the thermodynamics of the Saltzman model and the Lorenz system are also studied via an excess work, which is related to the necessary work to displace the system from the stationary state~\cite{velarde1994toward}. 
The Lorenz model may provide a basic dynamical and thermodynamic framework for considering more complex models.
}

  {In addition to the atmospheric motion, Lorenz equations can be applied in different areas; examples include DC motors~\cite{hemati1994strange}, chemical reaction systems~\cite{poland1993cooperative}, and a mechanical waterwheel model, known as the Malkus--Lorenz waterwheel~\cite{strogatz2018nonlinear,malkus1972non,lopez2022thermodynamic}. 
Recently, the thermodynamic efficiency of the Malkus--Lorenz waterwheel system was numerically and theoretically studied, and its maximum efficiency has been derived~\cite{lopez2022thermodynamic}.
It has also been found that the efficiency shows an abrupt drop at the point where the dynamics change from stationary to chaotic, and this can be applied to the more generic systems described by Lorenz equations other than the Malkus--Lorenz waterwheel system.
}

  {In this study, stimulated by~\cite{lopez2022thermodynamic}, the thermodynamic efficiency of atmospheric motion is studied based on Saltzman's model and Lorenz equations. Taking the mathematical similarities into consideration, it can be reasonably assumed that this efficiency is similar to  {the one} in~\cite{lopez2022thermodynamic}.
An analysis of the entropy generation rate will also bring out
the difference between the efficiencies of the atmospheric heat engines and those of conventional heat engines.
}

The remainder of this paper is structured as follows: A review of the setup of the Lorenz system is given in Sec.~\ref{secup}. Subsequently, the thermodynamic efficiency   {and the entropy generation rate} of atmospheric motion is calculated in Sec.~\ref{seceff}. To obtain efficiency, the heat flux and work necessary to maintain the atmospheric motion are determined. Then, in Sec.~\ref{secnum}, a numerical calculation of efficiency   {and entropy generation rate} is applied for one set of parameters, which can lead to chaotic dynamics with a relatively high Rayleigh number. Finally, we summarize the present study and discuss it in Sec.~\ref{secdis}.

\section{Lorenz system\label{secup}}
First, we review the setup of the Lorenz system~\cite{saltzman1962finite,lorenz1963deterministic}.
It was originally derived from the Oberbeck--Boussinesq approximation of fluid conservation equations, where the density variations are neglected, except when they give rise to a gravitational force~\cite{tritton2012physical}.
Let us consider the atmospheric motion between two parallel horizontal plates with distance $H$,
where each plate is externally heated to maintain its temperature such that the temperature difference $\Delta T$ between the plates is maintained constant.
Assuming that the atmospheric motion is restricted to a two-dimensional $x$-$z$ plane with vanishing velocity in $y$-direction, the Navier--Stokes equations and   {the} thermal convection equation in $x$-$z$ plane under the Oberbeck--Boussinesq approximation can be given as follows~\cite{saltzman1962finite}:
\begin{equation}
\begin{aligned}
\frac{\partial v_x}{\partial t} + v_x\frac{\partial v_x}{\partial x} + v_z\frac{\partial v_x}{\partial z} + \frac{\partial P}{\partial x} - \nu\nabla^2 v_x &=0,\\
\frac{\partial v_z}{\partial t} + v_x\frac{\partial v_z}{\partial x} + v_z\frac{\partial v_z}{\partial z} + \frac{\partial P}{\partial z} -g\alpha T - \nu\nabla^2 v_z &=0,
\end{aligned}\label{vfunc}
\end{equation}
\begin{equation}
\frac{\partial T}{\partial t} + v_x\frac{\partial T}{\partial x} + v_z\frac{\partial T}{\partial z} - \kappa \nabla^2 T=0.\label{Tfunc}
\end{equation}
Here, $v_x$ and $v_z$ are the velocities in $x$- and $z$-directions, respectively, satisfying the incompressibility condition $\frac{\partial v_x}{\partial x}+\frac{\partial v_z}{\partial z}=0$ in $x$-$z$ plane.
$T$ and $P$ are the temperature and pressure divided by density, respectively.
The constants $g$, $\alpha$, $\nu$, and $\kappa$ denote the   {gravitational acceleration}, coefficient of thermal expansion, kinematic viscosity, and   {coefficient of thermal diffusivity}, respectively.

Owing to the incompressibility condition, 
the stream function $\psi$ can be introduced for two-dimensional motion,
where the velocities in $x$- and $z$-directions are expressed as $v_x=-\frac{\partial \psi}{\partial z}$ and $v_z=\frac{\partial \psi}{\partial x}$, respectively.
We also introduce $\theta$, which is the nonlinear part of temperature $T$: 
\begin{equation}
T = T_h - \frac{\Delta T}{H}z + \theta.
\end{equation}
  {It} also denotes the departure of temperature from that in a state of vanishing convection~\cite{lorenz1963deterministic}, where $T_h\equiv T|_{z=0}$. 
Then, the governing equations (\ref{vfunc}) and (\ref{Tfunc}) can be rewritten in terms of $\psi$ and $\theta$ as~\cite{saltzman1962finite}
\begin{equation}
\frac{\partial}{\partial t}\nabla^2 \psi + \frac{\partial(\psi, \nabla^2 \psi)}{\partial(x,z)}-\nu\nabla^4\psi-g\alpha\frac{\partial\theta}{\partial z}=0,\label{psi}
\end{equation}
\begin{equation}
\frac{\partial}{\partial t}\theta + \frac{\partial(\psi, \theta)}{\partial(x,z)}-\frac{\Delta T}{H}\frac{\partial \psi}{\partial x} - \kappa\nabla^2\theta=0,\label{theta}
\end{equation}
where
\begin{equation}
\frac{\partial(a,b)}{\partial(x,z)}\equiv   {\frac{\partial a}{\partial x}\frac{\partial b}{\partial z} - \frac{\partial a}{\partial z}\frac{\partial b}{\partial x}}
\end{equation}
is the Jacobian operator

Using the method developed in~\cite{malkus1958finite} by scaling the length in $H$, time in $H^2/\kappa$, stream function in $\kappa$, and temperature in $(\kappa\nu)/(g\alpha H^3)$, a dimensionless version of Eqs.~\eqref{psi} and~\eqref{theta} can be given as
\begin{equation}
\frac{\partial}{\partial t^*}\nabla^{*2} \psi^* + \frac{\partial(\psi^*, \nabla^{*2} \psi^*)}{\partial(x^*,z^*)}-\sigma\nabla^{*4}\psi^*-\sigma\frac{\partial\theta^*}{\partial z^*}=0,\label{npsi}
\end{equation}
and 
\begin{equation}
\frac{\partial}{\partial t^*}\theta^* + \frac{\partial(\psi^*, \theta^*)}{\partial(x^*,z^*)}-R\frac{\partial \psi^*}{\partial x^*} - \nabla^{*2}\theta^*=0,\label{ntheta}
\end{equation}
with Prandtl number $\sigma \equiv \nu/\kappa$ and Rayleigh number $R \equiv (g\alpha H^3 \Delta T)/(\kappa\nu)$. Here, superscript `` $^*$ '' denotes the dimensionless replacement of the corresponding variable. \par

Naturally, for both the upper and lower boundaries, we impose
\begin{equation}
\theta^*|_{z^*=0,1} = 0\label{boundarytheta}
\end{equation}
because the temperatures of the two parallel horizontal plates are maintained constant. 
Moreover, we impose the following free boundary conditions on $\psi^*$ for tractability~\cite{lorenz1963deterministic}:
\begin{equation}
\begin{aligned}
\psi^*|_{z^*=0,1} &= 0,\\
\nabla^{*2}\psi^*|_{z^*=0,1} &= 0.
\end{aligned}\label{boundaryfree}
\end{equation}
By considering the boundary conditions Eqs. ~(\ref{boundarytheta}) and (\ref{boundaryfree}), modes $\cos (m\pi ax^*)\sin(n\pi z^*)$ and $\sin(m\pi ax^*)\sin(n\pi z^*)$ are available for $\theta^*$ and $\psi^*$, where $a^{-1}$ is the dimensionless wavelength in $x$-direction, and $m$ and $n$ are the wave numbers in $x$- and $z$-directions, respectively.   {These modes imply the repetitive convective cells with each height $H$ and length $Ha^{-1}$.}\par

Lorenz discovered that for a specific set of modes,
\begin{equation}
\begin{aligned}
a(1+a^2)^{-1}\psi^*&=X\sqrt{2}\sin(\pi ax^*)\sin(\pi z^*),\\
\pi R_c^{-1}\theta^*&=Y\sqrt{2}\cos(\pi ax^*)\sin(\pi z^*)-Z\sin(2\pi z^*),
\end{aligned}\label{lorenzmode}
\end{equation}
where 
\begin{equation}
R_c \equiv \pi^4(1+a^2)^3a^{-2},
\end{equation}
denotes the critical Rayleigh number,
$X$, $Y$ and $Z$ are governed by the following equations.
\begin{equation}
\begin{aligned}
\frac{dX}{d\tau}&=-\sigma X + \sigma Y,\\
\frac{dY}{d\tau}&=-XZ+rX-Y,\\
\frac{dZ}{d\tau}&=XY-bZ,
\end{aligned}\label{Lorenzeq}
\end{equation}
where $\tau \equiv \pi^2(1+a^2)t^*$, $r \equiv R/R_c$, and $b \equiv 4(1+a^2)^{-1}$~\cite{lorenz1963deterministic}. 
Equation~\eqref{Lorenzeq} is known as the Lorenz equation. There is one stable fixed point,
\begin{equation}
X=Y=Z=0,\label{fixed0}
\end{equation}
when $0 < r < 1$ and  {other} two fixed points
\begin{equation}
\begin{aligned}
X=Y=\pm\sqrt{b(r-1)}\ ,\ Z=r-1,
\end{aligned}\label{fixedpm}
\end{equation}
when $r > 1$ \cite{lorenz1963deterministic,strogatz2018nonlinear}. For $r > 1$, fixed point~\eqref{fixed0} is unstable, whereas fixed points~\eqref{fixedpm} are stable if $0 < \sigma < b+1$~\cite{lorenz1963deterministic,strogatz2018nonlinear}. However, if $\sigma > b+1$, fixed points~\eqref{fixedpm}   {are only stable when $r<\sigma(\sigma + b + 3)/(\sigma - b - 1)$} and   {become} unstable when $r > \sigma(\sigma + b + 3)/(\sigma - b - 1)$. The strange attractor called ``Lorenz attractor'' appears~\cite{strogatz2018nonlinear}. A   {supercritical} pitchfork bifurcation occurs at $r = 1$, and if $\sigma > b+1$, a subcritical Hopf bifurcation occurs at $r = \sigma(\sigma + b + 3)/(\sigma - b - 1)$~\cite{strogatz2018nonlinear}. It can be proven that $\sigma(\sigma + b + 3)/(\sigma - b - 1) > 1$ for any available $b$ when $\sigma > b+1$. \par

A mechanical waterwheel model is (partly) governed by the Lorenz equations known as the Malkus--Lorenz   {waterwheel}, in which water flows into a sloping wheel in a symmetry mode and leaks at a constant rate, causing the wheel to rotate against friction~\cite{strogatz2018nonlinear,malkus1972non,lopez2022thermodynamic}. This waterwheel model can be regarded as a generalized heat engine because it absorbs potential energy from the inflow and works against friction to maintain its rotation~\cite{lopez2022thermodynamic}. The thermodynamic efficiency is given by
\begin{equation}
\eta = \eta_{\rm max} \left(1-\frac{1}{r}\right),\label{wheeleff}
\end{equation}
when the system approaches stable fixed points~\eqref{fixedpm} and is bounded by the maximum efficiency $\eta_{\rm max}$ irrespective of whether the dynamic is chaotic ~\cite{lopez2022thermodynamic}. $r$ is the redefined Rayleigh number in the waterwheel model. \par

The thermodynamic efficiency of the Malkus--Lorenz {waterwheel} is related to the Rayleigh number, which, in the atmospheric motion situation, is related to the temperature difference $\Delta T$.   {As in the case of the Malkus--Lorenz waterwheel}, it is also necessary for the atmosphere to absorb energy, particularly heat energy, to maintain its motion. Thus, it is reasonable to regard atmospheric motion as a heat engine and the thermodynamic efficiency, which may also be bounded by maximum efficiency, can be expressed by the Rayleigh number $R$.   {In other words}, the efficiency may be similar to   {that expressed in} Eq.~\eqref{wheeleff} when   {the dynamics of the system} are stationary.
\par

Moreover, the thermodynamic efficiency of free convection is qualitatively given as:
\begin{equation}
\eta \sim \frac{g\alpha L}{C_p},\label{quaeta}
\end{equation}
where $L$ is the length scale of the convection system (for the situation considered in this study, $H$) and $C_p$ is the constant-pressure specific heat~\cite{tritton2012physical}. This expression may be related to maximum efficiency $\eta_{\rm max}$. \par

Thus, it is natural to assume that the efficiency studied here is a combination of Eq.~\eqref{wheeleff}  {and} Eq.~\eqref{quaeta}.

\section{Thermodynamic efficiency\label{seceff}}
\begin{figure}
\includegraphics[width=0.75\columnwidth]{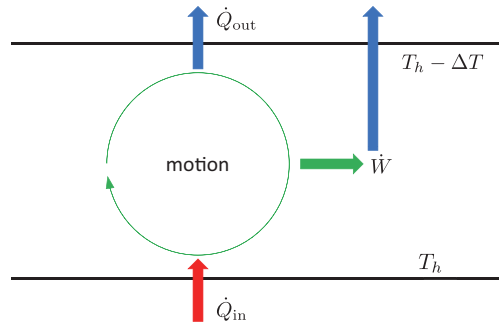}
\caption{Illustration of the atmospheric motion when it is regarded as a heat engine. Heat is absorbed at the hot boundary ($z=0$ with temperature $T_h$) as $\dot{Q}_{\rm in}$ (red bold arrow), and dissipated at the cold boundary ($z=H$ with temperature $T_h-\Delta T$) as $\dot{Q}_{\rm out}$ (blue bold arrows). A part of the heat absorbed is converted to the necessary work $\dot{W}$ (green bold arrow) to maintain the motion (green ring arrow). However, this work is usually dissipated again~\cite{tritton2012physical}.\label{figsketch}}
\end{figure}
Figure~\ref{figsketch} shows a schematic of atmospheric motion. The heat $\dot{Q}_{\rm in}$ is absorbed per unit time at the hot boundary $z=0$, and a part of it is converted to work per unit time $\dot{W}$ to maintain the motion. The heat $\dot{Q}_{\rm out}$ is dissipated per unit time at the cold boundary $z=H$. The  efficiency can be defined as:
\begin{equation}
\eta \equiv \frac{\dot{W}}{\dot{Q}_{\rm in}}.\label{effdif}
\end{equation}
  {It must be noted} that $\dot{W}$ is usually dissipated again, such that $\dot{Q}_{\rm in} \approx \dot{Q}_{\rm out}$ \cite{tritton2012physical}. \par

Moreover, it is necessary to revert to the governing equations (\ref{vfunc}) and (\ref{Tfunc}) before calculating heat absorbed per unit time $\dot{Q}_{\rm in}$, which is related to heat flux $\mathbf{j}$, and work per unit time $\dot{W}$. 

\subsection{Heat flux}
It is easy to obtain the heat flux $\mathbf{j}=(j_x, j_z)$ that satisfies
\begin{equation}
\frac{\partial T}{\partial t} + \nabla\cdot\mathbf{j}=0
\end{equation}
from Eq.~\eqref{Tfunc} with   {the} incompressibility condition in $x$-$z$ plane. Here, $\mathbf{j}$ can be chosen as
\begin{equation}
\mathbf{j} = \mathbf{v}T-\kappa\nabla T,
\end{equation}
which is the total heat flux including convection term $\mathbf{v}T$ where $\mathbf{v}=(v_x, v_z)$ and conduction term $-\kappa\nabla T$~\cite{saltzman1962finite}.
In particular, in $z$-direction, we have
\begin{equation}
j_z = v_z T - \kappa\frac{\partial T}{\partial z} = \frac{\partial \psi}{\partial x}T - \kappa\frac{\partial T}{\partial z}.
\end{equation}%\par
$j_z$ at the boundary is related to the absorbed and dissipated heat, respectively. By applying the mode in Eq.~\eqref{lorenzmode} and integrating $j_z$ at both the hot boundary $(\partial V)_h = \{(x,z)|z=0, 0<x<Ha^{-1}\}$ and the cold boundary $(\partial V)_c = \{(x,z)|z=H, 0<x<Ha^{-1}\}$, the absorbed heat per unit time can be calculated as follows:
\begin{equation}
\dot{Q}_{\rm in} = \rho C_p\int_{(\partial V)_h}j_z dx=\rho C_p\frac{\kappa^2\nu}{ga\alpha H^3}(R + 2R_c Z),\label{Qin}
\end{equation}
whereas the dissipated heat per unit time is
\begin{equation}
\dot{Q}_{\rm out} = \rho C_p\int_{(\partial V)_c}j_z dx=\rho C_p\frac{\kappa^2\nu}{ga\alpha H^3}(R + 2R_c Z).\label{Qout}
\end{equation}
Here, $\dot{Q}_{\rm in} = \dot{Q}_{\rm out}$, which implies that the work per unit time $\dot{W}$ to maintain the motion is finally dissipated.

\subsection{Work to maintain the atmospheric motion}
The kinetic energy per unit mass is defined as follows:
\begin{equation}
e_K \equiv \frac{1}{2}\left(v_x^2 + v_z^2\right).
\end{equation}
Combined with Eq.~\eqref{vfunc}, the energy conservation equation can be expressed as follows:
\begin{equation}
\frac{\partial e_K}{\partial t} = \nu \mathbf{v}\cdot(\nabla^2\mathbf{v}) + g\alpha v_z T - \nabla\cdot[(e_K+P)\mathbf{v}].\label{funce}
\end{equation}
Here, $\nu\mathbf{v}\cdot(\nabla^2\mathbf{v})$ is related to the necessary work per unit time to maintain the motion, $-\nabla\cdot[(e_K+P)\mathbf{v}]$ is the convection term with no energy input and output, as $[(e_K+P)v_z]|_{z=0,H} = 0$ when considering the boundary conditions~\eqref{boundarytheta} and~\eqref{boundaryfree}, and $g\alpha v_z T$ can be expressed by an ``available potential energy'' $-(g\alpha H\theta^2)/(2\Delta T)$ when considering the average over the entire fluid~\cite{saltzman1962finite}.   {It should be noted that the mechanical work is assumed   {to be} only dissipated as the frictional heat. Although other factors, such as phase changes in the water cycle, also play important roles in a climate system \cite{lembo2019thediato, pauluis2002Aentropy, pauluis2002Bentropy}, they do not appear in the governing equations~\eqref{vfunc} and ~\eqref{Tfunc}.}\par

Considering the mode in Eq.~\eqref{lorenzmode}, the necessary work per unit time $\dot{W}$ to maintain the motion can be calculated by integrating $-\rho\nu\mathbf{v}\cdot\left(\nabla^2\mathbf{v}\right)$ over the entire motion space $V = \{(x,z)|0 < x < Ha^{-1}, 0 < z < H\}$ as
\begin{equation}
\dot{W} = -\rho\nu\int_{V}\mathbf{v}\cdot\left(\nabla^2\mathbf{v}\right)dV = \frac{2\kappa^2\nu\rho R_c}{abH^2}X^2.\label{W}
\end{equation}

\subsection{Efficiency}
It appears that by using $\dot{Q}_{\rm in}$ in Eq.~(\ref{Qin}) and $\dot{W}$ in Eq.~(\ref{W}), the thermodynamic efficiency can be calculated directly using Eq.~\eqref{effdif}. However, these dynamics may become chaotic under certain parameter sets. Thus, it is necessary to determine whether the dynamics are stationary or chaotic to obtain efficiency. \par

The case is straightforward if the dynamics approach a fixed point. Using Eqs.~\eqref{Qin} and~\eqref{W}, the efficiency can be expressed as
\begin{equation}
  {\eta = \frac{\dot{W}}{\dot{Q}_{\rm in}} = \frac{g\alpha H}{C_p}\frac{2 X^2}{b(R/R_c+2 Z)}}\quad \text{(stationary dynamics)}.
\end{equation}
The fixed point~\eqref{fixed0} is approached if $0 < R/R_c < 1$, whereas the fixed points~\eqref{fixedpm} is approached if $R/R_c > 1$ and $0<\sigma<b+1$, or if $1<R/R_c<\sigma(\sigma+b+3)/(\sigma-b-1)$ and $\sigma>b+1$. Thus, the efficiency can be calculated as:
\begin{equation}
\eta=\begin{cases}
0& \text{(stable fixed point~\eqref{fixed0})},\\
\frac{g\alpha H}{C_p}  {\frac{2(R/R_c-1)}{3(R/R_c)-2}}&\text{(stable fixed points~\eqref{fixedpm})}.
\end{cases}\label{etastable}
\end{equation}
For chaotic dynamics, if $R/R_c > \sigma(\sigma+b+3)/(\sigma-b-1)$ and $\sigma > b+1$, it cannot be derived as a closed expression for efficiency because the dynamics evolve through the chaotic attractor~\cite{lopez2022thermodynamic}. Nevertheless, $\dot{W}$ and $\dot{Q}_{\rm in}$ can be averaged because the limit set is densely covered by a particular trajectory~\cite{lopez2022thermodynamic}. The average of a function $f$ is defined as follows:
\begin{equation}
\left<f\right>\equiv\lim_{\tau\to +\infty}\frac{1}{\tau}\int_{0}^{\tau}f dt.
\end{equation}
Thus, the efficiency can be expressed as
\begin{equation}
\begin{aligned}
\eta = \frac{\left<\dot{W}\right>}{\left<\dot{Q}_{\rm in}\right>} &=   {\frac{g\alpha H}{C_p}\frac{2 \left<X^2\right>}{b(R/R_c+2 \left<Z\right>)}}\\
&\qquad\text{(chaotic dynamics)}.
\end{aligned}\label{etachaotic}
\end{equation}

\subsection{  {Entropy generation rate}}
  {The present system can be considered to be in a steady state 
with $\dot{Q}_{\rm in} = \dot{Q}_{\rm out}$
in a statistical sense~\cite{ozawa2003second}, no matter whether the dynamics is stationary or chaotic.
 Therefore, the entropy generation rate of the total system should be equal to the entropy increase rate in the surrounding environment~\cite{ozawa2003second}:
\begin{equation}
    \dot{S} \equiv \frac{\dot{Q}_{\rm out}}{T_h - \Delta T} - \frac{\dot{Q}_{\rm in}}{T_h}.\label{EPG}
\end{equation}
By time averaging Eq.~\eqref{EPG}, we have
\begin{equation}
    \begin{aligned}
         \left<\dot{S}\right> &\simeq \frac{\rho C_p\kappa^2\nu}{ga\alpha H^3}(R + 2R_c \left<Z\right>)\frac{\Delta T}{T_h}\\
         &=\frac{\rho C_p\kappa^3\nu^2R_c^2}{aT_h(g\alpha H^3)^2}\left(\frac{R}{R_c}\right)\left(\frac{R}{R_c} + 2 \left<Z\right>\right)\\
         &=\varepsilon \left(\frac{R}{R_c}\right)\left(\frac{R}{R_c} + 2 \left<Z\right>\right),
    \end{aligned}\label{entropy}
\end{equation}
for both stationary and chaotic dynamics,
where we have defined the  {parameter $\varepsilon$} with a dimension of the entropy generation rate as
\begin{equation}
\varepsilon \equiv \frac{\rho C_p\kappa^3\nu^2R_c^2}{aT_h(g\alpha H^3)^2}.
\end{equation}
Here, in addition to $\dot{Q}_{\rm in} = \dot{Q}_{\rm out}$, we used Eq.~\eqref{Qin} and the definition of Rayleigh number $R$, and assumed 
\begin{equation}
    \frac{1}{T_h - \Delta T} \simeq \frac{1}{T_h}\left(1 + \frac{\Delta T}{T_h}\right),
\end{equation}
due to the Oberbeck--Boussinesq approximation, which claims $\Delta T \ll T_h$.
}
  {Especially, when the dynamics is stationary, Eq.~\eqref{entropy} becomes
\begin{equation}
    \left<\dot{S}\right> = \varepsilon\left(\frac{R}{R_c}\right)^2, \label{eq.S_gen_1} 
\end{equation}
when approaching to the stable fixed point in Eq.~\eqref{fixed0}, or
\begin{equation}
    \left<\dot{S}\right> = \varepsilon\left(\frac{R}{R_c}\right)\left[3\left(\frac{R}{R_c}\right) - 2\right], \label{eq.S_gen_2} 
\end{equation}
when approaching to the stable fixed point in~Eq.~\eqref{fixedpm}.
}

\section{Numerical calculation\label{secnum}}
\subsection{  {Efficiency}}
  {We easily find that the efficiency in Eq.~\eqref{etastable} for the stationary dynamics is bounded from above by $(2g\alpha H)/(3C_p)$, for example, for $0 < \sigma < b+1$.} Although the expression of the efficiency cannot be derived as a closed one under chaotic dynamics,   {we may expect} that the efficiency shares the same upper bound, $(2g\alpha H)/(3C_p)$, as in the case of stationary dynamics,   {which will be shown in Appendix~\ref{drop_Z}.}\par

\begin{figure}
\includegraphics[width=0.95\columnwidth]{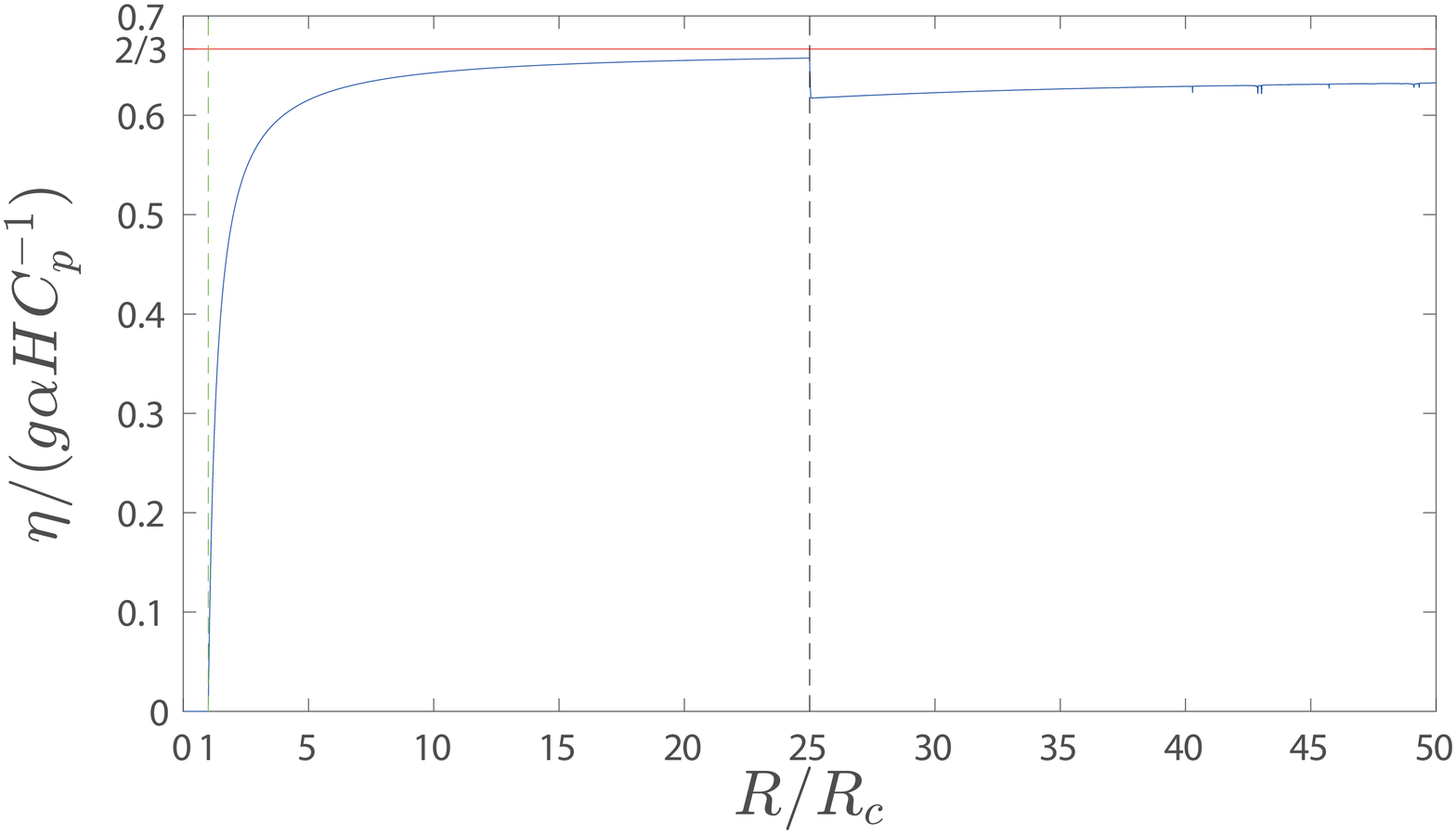}
\caption{$\eta/(g\alpha HC_p^{-1})$ vs. $R/R_c$. Parameters for the corresponding Lorenz equations Eq.~\eqref{Lorenzeq} are chosen as $b=2$ and $\sigma = 5$. Pitchfork bifurcation occurs at $R/R_c = 1$ (green dashed line), and subcritical Hopf bifurcation occurs at $R/R_c=\sigma(\sigma+b+3)/(\sigma-b-1) = 25$ (black dashed line)~\cite{strogatz2018nonlinear,malkus1972non,lopez2022thermodynamic}. $\eta/(g\alpha HC_p^{-1})$ (blue solid line) is bounded by $2/3$ (red solid line) irrespective of whether the dynamics are stable ($R/R_c < 25$) or chaotic ($R/R_c > 25$). However, there is an abrupt drop at $R/R_c = 25$, where the dynamics start to become chaotic.
  {We used the fourth Runge-Kutta method with $1.2\times 10^{6}$ steps and each time step width $\Delta \tau = 0.01$ in Eq.~\eqref{Lorenzeq}. The last $1\times 10^{6}$ steps are used to calculate the time average.}
~\label{figeta}}
\end{figure}

Figure~\ref{figeta} shows the numerical calculation results of the non-dimensionalized efficiency $\eta/(g\alpha HC_p^{-1})$, plotted as a blue solid line. The necessary parameters were chosen as $b=2$ and $\sigma=5$, making it possible for the dynamics to become chaotic. The two dashed lines in the figure represent the pitchfork bifurcation point at $R/R_c = 1$ (green dashed line) and subcritical Hopf bifurcation point $R/R_c = \sigma(\sigma+b+3)/(\sigma-b-1) = 25$ (black dashed line). The stable fixed point~\eqref{fixed0} is approached when $0<R/R_c<1$ and the efficiency vanishes, whereas a stable fixed point~\eqref{fixedpm} is approached when $1<R/R_c<25$. The efficiency is described by Eq.~\eqref{etastable} in stationary dynamics and is calculated using Eq.~\eqref{etachaotic}, when in chaotic dynamics. \par

Except for the discontinuous point at $R/R_c = 25$, the efficiency increases as $R/R_c$ increases when $R/R_c > 1$. Because $R \propto \Delta T$, the efficiency increases as the temperature difference increases. This type of behavior is commonly observed in the efficiency of heat engines operating between two heat reservoirs at different temperatures, such as the Carnot heat engine~\cite{lopez2022thermodynamic}. 
It is crucial to recognize the increasing complexity due to various additional factors in considering the large-scale climate regime. 
 {
For example, moisture plays an important role in the efficiency of the climate system~\cite{lembo2019thediato,pauluis2002Aentropy,pauluis2002Bentropy}, which is not captured by the model presented in this study.}\par

Moreover, regardless of whether stationary or chaotic, $\eta/(g\alpha HC_p^{-1})$ is bounded by $2/3$ (red solid line), implying that   {(see Appendix~\ref{drop_Z} for the derivation)}
\begin{equation}
\eta < \frac{2g\alpha H}{3C_p}.\label{etabound}
\end{equation}
Although the value of the upper bound $(2g\alpha H)/(3C_p)$ is not explicitly mentioned, there should be no concern regarding the forbidden situation $\eta > 1$ because $H$ is limited. Otherwise, the Oberbeck-Boussinesq approximation does not hold because density differences cannot be ignored~\cite{tritton2012physical}. In fact, $g\alpha H/C_p$ should be very small according to the qualitative description in~\cite{tritton2012physical}. \par

An interesting drop in efficiency occurs at $R/R_c = 25$ because $\left<Z\right>$ drops when the dynamics become chaotic (see also~\cite{lopez2022thermodynamic}). An explanation of the  {decrease} in $\left<Z\right>$ is   {provided} in the Appendix \ref{drop_Z}.

\subsection{  {Entropy generation rate}}
\begin{figure}
    \includegraphics[width=0.95\columnwidth]{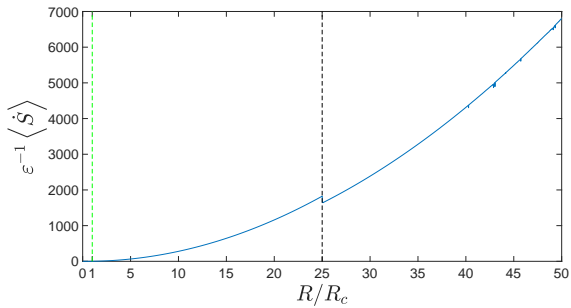}
    \caption{  {$\varepsilon^{-1}\left<\dot{S}\right>$ vs. $R/R_c$. Parameters for the corresponding Lorenz equations Eq.~\eqref{Lorenzeq} are chosen as $b=2$ and $\sigma = 5$, which are the same as those in Fig.~\ref{figeta}. We used the fourth Runge-Kutta method with $1.2\times 10^{6}$ steps and each time step width $\Delta \tau = 0.01$ in Eq.~\eqref{Lorenzeq}. The last $1\times 10^{6}$ steps are used to calculate the time average.}\label{entropy_graph}}
\end{figure}
  {Figure \ref{entropy_graph} shows the numerical results of the entropy generation rate $\varepsilon^{-1}\left<\dot{S}\right>$. The parameters in the Lorenz equations \eqref{Lorenzeq} are chosen as $b=2$ and $\sigma = 5$.} 
    
  {Basically, no matter whether the dynamics are stable or chaotic, $\left<\dot{S}\right>$ in Eq.~\eqref{entropy} monotonically increases 
as $R/R_c$ increases; especially, 
$\left<\dot{S}\right>$ in Eq.~\eqref{eq.S_gen_2}
for the stationary dynamics behaves as
\begin{equation}
    \left<\dot{S}\right> \sim \left(\frac{R}{R_c}\right)^2.
\end{equation}
The exception is the Hopf bifurcation point $R/R_c = 25$,
where the abrupt drop of $\left<\dot{S}\right>$ is evident.}

  {
Upon comparing Figs.~\ref{figeta} and~\ref{entropy_graph}, it is evident that, for both stationary and chaotic dynamics,
the efficiency increases despite the increase of the entropy generation rate. This is consistent with the results of~\cite{lopez2022thermodynamic}, but is clearly different from the conventional thermodynamic efficiency of heat engines, where the increase of irreversibility generally reduces the efficiency.
This is attributed to the fact that the 
generated work in the atmospheric heat engine is eventually dissipated as heat into the cold heat reservoir.
}

\section{Summary and Discussion\label{secdis}}
In this study, the thermodynamic efficiency of the atmospheric motion governed by the Lorenz system was determined and calculated. It is shown in Sec.~\ref{seceff} that the heat input from the hot boundary is equal to the heat output to the cold boundary because the work, which is converted from a part of the heat input, dissipates again~\cite{tritton2012physical}. \par

An upper bound of the efficiency exists when the Rayleigh number $R$ varies, which is similar to the results in~\cite{lopez2022thermodynamic}, regardless of whether the dynamics are chaotic   {or not}. The upper  {bound} has the same structure as that given qualitatively in~\cite{tritton2012physical}. \par

For the parameter sets that can cause the dynamics to be chaotic, an abrupt drop in efficiency occurs at Hopf bifurcation. This drop, which is   {because of} the discontinuity of $\left<Z\right>$, is also discovered in the Malkus--Lorenz waterwheel system~\cite{lopez2022thermodynamic}. \par

{As for the entropy generation rate, the similar drop as the efficiency also happens at the Hopf bifurcation point.}

 {It is in our expect to find these similarities when compared with the Malkus-Lorenz waterwheel system because they are commonly described by the Lorenz equations.}
However, {the efficiency and the entropy generation rate in this study are not exactly the same as those in~\cite{lopez2022thermodynamic}.} 
Both the ``Rayleigh number'' and ``Prandtl number'' are variable in the Malkus--Lorenz   {waterwheel} system by adjusting friction rate ``$\nu$'' and water inflow mode, but the Prandtl number is fixed here for the specific fluid, and Rayleigh number can be changed by only adjusting $\Delta T$. 
Moreover, the length scale $H$ is limited   {because of} the Oberbeck--Boussinesq approximation~\cite{tritton2012physical}.
Despite these differences, such similarities are interesting.\par
  {Notably, as a simplified model, the present model governed by the Lorenz system has certain limitations. Its application in the large-scale climate modeling is challenging because of the Oberbeck--Boussinesq approximation. Moreover, the frictional dissipation in the Saltzman model accounts only for a small fraction of the entropy generation rate, where the latent heat transport given by moisture and its phase changes are not taken into consideration~\cite{lembo2019thediato,pauluis2002Aentropy,pauluis2002Bentropy}. Further, in addition to the vertical convection considered in the Saltzman model, horizontal heat transportation, which is dominant at the midlatitude, also plays an important role~\cite{lucarini2011new, juckes2000linear}. }\par
  {Despite these limitations, the present study makes significant contributions} to the understanding of the complex climate system from a thermodynamics perspective.
   {Further, we intend to add other factors to this simple model and find their properties in both dynamics and thermodynamics. We expect that similar behaviors, such as the discontinuities of efficiency and entropy generation rate at certain bifurcation points, still exist.}

\begin{acknowledgements}
This work was supported by JSPS KAKENHI Grant Numbers 19K03651 and 22K03450.
\end{acknowledgements}

\appendix
\section{ {Decrease} of $\left<Z\right>$   {and derivation of Eq.~\eqref{etabound}}}\label{drop_Z}
The average $\left<\cdot\right>$ is a linear operator, and for a limited function $f$, it can be proven that
\begin{equation}
\left<\frac{df}{dt}\right> = 0,\label{aved}
\end{equation}
\begin{equation}
\left<f^2\right> \ge 0
\end{equation}
and
\begin{equation}
\left<f^2\right> \ge \left<f\right>^2\label{avege}
\end{equation}
with equality if and only if $f$ is a constant. \par
Note that the variables in the Lorenz equations in Eq.~\eqref{Lorenzeq} are limited because they approach a stable fixed point or evolve through a chaotic attractor~\cite{lopez2022thermodynamic}. Applying Eq.~\eqref{aved} to the third equation in Eq.~\eqref{Lorenzeq} gives 
\begin{equation}
\left<XY\right> = b\left<Z\right>.\label{XY_bZ}
\end{equation}
Multiplying $X$ on both sides of the first equation in Eq.~\eqref{Lorenzeq}, and applying Eq.~\eqref{aved}, we obtain
\begin{equation}
\left<XY\right> = \left<X^2\right>.\label{XY_X2}
\end{equation}
  {From Eqs.~\eqref{XY_bZ}  {and \eqref{XY_X2}}, we have
\begin{equation}
b\left<Z\right>=\left<X^2\right>.\label{bZ_X2}
\end{equation}
Therefore, from Eq.~\eqref{bZ_X2}, the efficiency  Eq.~\eqref{etachaotic} can be re-expressed as:
\begin{equation}
\eta=\frac{g\alpha H}{C_p}\left(1-\frac{R/R_c}{R/R_c+2 \left<Z\right>}\right).\label{reeta}
\end{equation}}
The decrease in $\eta$ at $R/R_c = \sigma(\sigma + b + 3)/(\sigma - b - 1)$ indicates  {a decrease} in $\left<Z\right>$ at the same point. \par
Multiplying $Y$ on both sides of the second equation and $Z$ on both sides of the third equation of Eq.~\eqref{Lorenzeq}, followed by applying Eq.~\eqref{aved}, gives the following
\begin{equation}
\begin{aligned}
\left<XYZ\right> &= r\left<XY\right> - \left<Y^2\right>\\
&=rb\left<Z\right> - \left<Y^2\right>,
\end{aligned}
\end{equation}
and
\begin{equation}
\begin{aligned}
\left<XYZ\right> = b\left<Z^2\right>.
\end{aligned}
\end{equation}
Equation~\eqref{avege} shows $\left<Z^2\right> \ge \left<Z\right>^2$ with equality if and only if the dynamics are stationary. Thus,
\begin{equation}
\left<XYZ\right> \ge b\left<Z^2\right>.\label{lowerxyz}
\end{equation}
Consider the following relationship:
\begin{equation}
\begin{aligned}
0 = \left<X - Y\right>^2 &\le \left<(X-Y)^2\right>\\
&= \left<X^2\right> - 2\left<XY\right> + \left<Y^2\right>\\
&= -b\left<Z\right> + \left<Y^2\right>,
\end{aligned}
\end{equation}
with equality if and only if the dynamics are stationary, which implies that
\begin{equation}
\left<XYZ\right> \le (r-1)b\left<Z\right>.\label{upperxyz}
\end{equation}
With Eq.~\eqref{lowerxyz} and Eq.~\eqref{upperxyz}, we obtain:
\begin{equation}
\left<Z\right>^2 \le (r-1)\left<Z\right>.
\end{equation}
For $r = R/R_c \ge 1$,
\begin{equation}
0 \le \left<Z\right> \le r-1.
\end{equation}
  {It is worth noting} that $r-1$ is the value of $Z$ at fixed points~\eqref{fixedpm}, $\left<Z\right> = r-1$ when the dynamics are stationary, and $\left<Z\right> < r-1$ when the dynamics are chaotic.
  %{Thus, an abrupt drop in $Z$ occurs at $R/R_c = \sigma(\sigma + b + 3)/(\sigma - b - 1)$, which is the point at which   {the dynamics go from stationary to chaotic leading} to a drop in $\eta$ at the same point (see Fig.~\ref{figeta})}.
   {This will lead to the drop in both efficiency and entropy generation rate, which is consistent with the numerical calculations (see Fig.~\ref{figeta} and \ref{entropy_graph}). Moreover, the abrupt drop in the numerical calculations implies the discontinuity of $\left<Z\right>$ at $R/R_c = \sigma(\sigma + b + 3)/(\sigma - b - 1)$, which is the subcritical Hopf bifurcation point.}
  {Finally, by applying $\left<Z\right> < r-1$ for chaotic dynamics to Eq.~\eqref{reeta}, we have
\begin{equation}
\eta <\frac{g\alpha H}{C_p}  {\frac{2(R/R_c-1)}{3(R/R_c)-2}}.
\end{equation}
  {Thus}, the efficiency for the chaotic dynamics has been proved to be lower than the efficiency for the stationary dynamics in Eq.~\eqref{etastable} corresponding to Eq.~\eqref{fixedpm}, and we have thus derived Eq.~\eqref{etabound}.
}

%\begin{acknowledgments}
%\end{acknowledgments}
%\bibliography{ref_YI}


\end{document}